\documentclass[conference]{IEEEtran}
\IEEEoverridecommandlockouts
\usepackage{cite}
\usepackage{amsmath,amssymb,amsfonts}
\usepackage{algorithm2e}
\usepackage{graphicx}
\usepackage{textcomp}
\usepackage{xcolor}
\def\BibTeX{{\rm B\kern-.05em{\sc i\kern-.025em b}\kern-.08em
    T\kern-.1667em\lower.7ex\hbox{E}\kern-.125emX}}
\begin{document}

\title{Clustered Latent Dirichlet Allocation for Scientific Discovery
\thanks{This work has been supported by HPCC Systems, LexisNexis Risk Solutions, RELX Group, Elsevier Scopus, Department of Education GAANN award P200A150310, and National Science Foundation awards \#1228312 and \#1405767.}
}

\author{

\IEEEauthorblockN{1\textsuperscript{st} Christopher Gropp}
\IEEEauthorblockA{\textit{Oak Ridge National Laboratory}\\
Knoxville, TN, United States \\
groppcw@ornl.gov}
\and
\IEEEauthorblockN{2\textsuperscript{nd} Alexander Herzog}
\IEEEauthorblockA{\textit{Clemson University}\\
Clemson, SC, United States \\
aherzog@clemson.edu}
\and
\IEEEauthorblockN{3\textsuperscript{rd} Ilya Safro}
\IEEEauthorblockA{\textit{Clemson University}\\
Clemson, SC, United States \\
isafro@clemson.edu}
\and
\IEEEauthorblockN{4\textsuperscript{th} Paul W. Wilson}
\IEEEauthorblockA{\textit{Clemson University}\\
Clemson, SC, United States \\
pww@clemson.edu}
\and
\IEEEauthorblockN{5\textsuperscript{th} Amy W. Apon}
\IEEEauthorblockA{\textit{Clemson University}\\
Clemson, SC, United States \\
aapon@clemson.edu}

}

\maketitle

\begin{abstract}
Topic modeling, a method for extracting the underlying themes from a collection of documents, is an increasingly important component of the design of intelligent systems enabling the sense-making of highly dynamic and diverse streams of text data. Traditional methods such as Dynamic Topic Modeling (DTM) do not lend themselves well to direct parallelization because of dependencies from one time step to another. In this paper, we introduce and empirically analyze Clustered Latent Dirichlet Allocation (CLDA), a method for extracting dynamic latent topics from a collection of documents. Our approach is based on data decomposition in which the data is partitioned into segments, followed by topic modeling on the individual segments. The resulting local models are then combined into a global solution using clustering. The decomposition and resulting parallelization leads to very fast runtime even on very large datasets. Our approach furthermore provides insight into how the composition of topics changes over time and can also be applied using other data partitioning strategies over any discrete features of the data, such as geographic features or classes of users. In this paper CLDA is applied successfully to seventeen years of NIPS conference papers (2,484 documents and 3,280,697 words), seventeen years of computer science journal abstracts (533,588 documents and 46,446,184 words), and to forty years of the PubMed corpus (4,025,976 documents and 386,847,695 words). On the PubMed corpus, we demonstrate the versatility of CLDA by segmenting the data by both time and by journal.
\end{abstract}


\section*{Introduction}
Topic modeling, a method for extracting the underlying themes from a collection of documents, is an increasingly important component of the design of intelligent systems enabling the sense-making of highly dynamic and diverse streams of text data \cite{multimedia-news-summarization,latent-community-topic-analysis}. One of the most common models used in practice is Latent Dirichlet Allocation (LDA) \cite{LDA}.  With LDA, documents are assumed to be randomly generated from one or more topics, each of which is a distribution of words.  The topics are viewed as latent variables, and LDA executes by inferring the topics from the documents via a Dirichlet process. The algorithm repeatedly 
samples the documents and modifies the topics to better fit them until reaching a specified convergence. LDA has a number of assumptions, including that both words and documents are unordered and that all documents are generated in the same timeframe. 

Of interest in streaming Big Data analytics is the modeling of topics found in a dynamic stream of data, for example, a social media data stream that changes quickly in time \cite{Cataldi-personal-topic-detection}, a massive collection of publications that have been produced over long time steps \cite{DTM}, or discretized sensor data \cite{Farrahi-routines}. Dynamic Topic Modeling (DTM) \cite{DTM} relaxes the assumption of LDA that all documents are generated in the same time step. The corpus is divided into a sequence of time steps.  A fixed count of topics is estimated by the DTM method and the set of topics spans all time steps, but the
most important words extracted in each time step for a particular topic are allowed to change through time.  The estimation of the topics in a time step is dependent on the estimation from the previous time step.   DTM enables observation of how the language of a topic changes over time, and also how well represented a topic is at any given point in time. 

The DTM algorithm is, in general, much slower than that of LDA.  Because of the dependencies, the application of the DTM algorithm and implementation developed by Blei and Gerrish \cite{DTMcode} is limited to modest-sized datasets.  DTM, while mathematically elegant, does not lend itself well to direct parallelization because of dependencies from one time step to another, though a few recent attempts have made progress in this area~\cite{ParallelDTM}. \\

\noindent {\bf Our Contribution:}
We propose an alternative approach to modeling topic dynamics, namely, Clustered Latent Dirichlet Allocation (CLDA). Our approach is based on data decomposition in which the data are partitioned into segments, followed by topic modeling on the individual segments. The resulting local models are then combined into a global solution using clustering. We implement this approach using a fast, parallel algorithm for LDA \cite{PLDAplus} for inferring the local models, and k-means clustering \cite{parallelk-means} for combining the results. Our approach has several advantages. The decomposition and resulting parallelization leads to very fast runtime even on very large datasets. The clustering of local topics provides additional insights into topic dynamics; for example, a topic can emerge, die off, or split into multiple local topics, while preserving a global view of topic representation. We compare CLDA to DTM and LDA using metrics of runtime performance, perplexity, and the similarity of topics produced.  We report strong results on all of these metrics. The implementation of CLDA is available at \cite{CLDAimplementation}.\\

An overview of the steps of CLDA is as follows. First, the data are discretized into segments using time steps or other criteria such as geographic location or data source.
Each of these segments is a sub-corpus that is used as the input to a separate run of PLDA+ \cite{PLDAplus}, a highly parallelized implementation of LDA.  Since processing each segment is an independent task, the runs of PLDA+ on these several segments can also be performed in parallel.  The output for this step is a set of topics for every segment. The full list of these topics is passed to a parallelized implementation of k-means clustering \cite{parallelk-means}, producing a set of topics representative of the full set. Because each step of the method uses highly parallelized code, and because the estimation of local topics can be further independently parallelized, CLDA is highly scalable and fast on even very large datasets.    As a result of clustering, each original topic is a member of a particular global topic cluster, which is represented by its centroid. 

Mathematical analysis of CLDA as compared to DTM is intractable. In this paper we empirically compare CLDA to the original DTM implementation and PLDA+ with respect to runtime, the quality of topics, and the similarity of the topics that are produced. We apply CLDA successfully to seventeen years of NIPS conference papers (2,484 documents and 3,280,697 words), seventeen years of computer science journal abstracts (533,560 documents and 40,002,197 words), and to forty years of the PubMed corpus (4,025,978 documents and 273,853,980 words).  Our experiments show that CLDA executes two orders of magnitude faster than the original DTM implementation and, as a result, can be practically used for processing  much larger datasets than is possible with DTM.

We evaluate CLDA on the quality of the topic models that are produced.  Perplexity is traditionally used as the measure of the quality of a topic model, and describes how close the model fit is to a held-out dataset. Our results show that the perplexity of CLDA is comparable to that of DTM and LDA on the same dataset and using the same global topic parameter.  

A key challenge we face in this research is the identification of a robust measure for comparing the similarity of two different topic models.  Perplexity does not measure how similar the topics are to each other for different models.  For the goal of comparing the actual topics produced by the DTM and CLDA modeling approaches we need an additional metric. For this last evaluation we choose to apply set-based measures, including the S{\o}rensen-Dice coefficient and the Jaccard index, to compare the top most frequently occurring subset of words of the topic.  Our results show that topics generated by CLDA are similar, using set-based metrics, to those generated by DTM and LDA. 

CLDA has a number of promising characteristics.
Unlike DTM, which requires that all time steps have the same number of topics, CLDA allows for the birth and death of topics between time steps. CLDA also facilitates in-depth exploration of topics within time segments without sacrificing information about global trends. CLDA scales favorably with the number of processors, the size of the document corpus, and the number of topics across even very large datasets. 
In addition, analysis can be performed on the composition of each global topic in each segment, allowing a better fit for individual time steps than DTM.  Matching the original topic mixtures to their representative centroids also enables comparison across time of global topics and analysis of how the global topics change over time.  



\section*{Background and Related Work}\label{sec:related_work}
\subsection*{Topic Modeling}

Clustered Latent Dirichlet Allocation is presented as an extension of LDA for analyzing large corpora that can be partitioned into segments~\cite{LDA}.  LDA incorporates a number of assumptions.  These include assumptions that words are unordered, topics are distributions of words, and multiple topics can contribute to a document that is a mixture of topics. Additionally, there is an assumption that the prior distribution of each topic is a Dirichlet distribution, which distinguishes it from more generalized methods. Only the output of the model (i.e., documents)
can be observed directly. The topics and topic mixtures are latent variables that must be inferred. 

With the LDA method, documents are assumed to be generated by sampling topics from their topic mixture, and sampling words from those topics, and repeating this process to generate all words in the document.
Topics are randomly seeded, and then iteration proceeds using Bayesian inference.  During each iteration, LDA compares each document with the topic and updates the topics for the next iteration, which continues until a stopping criteria is met.  Convergence can be measured either by change in the inferred parameters, or by another objective metric of the model such as the likelihood of producing the input set \cite{LDA}. Formally, for a mixture of $K$ latent topics, where topic $k$ is a multinomial distribution $\phi_k$ over a $W$-word vocabulary, for any document $D_j$, its topic mixture $\theta_j$ is a probability distribution drawn from a Dirichlet prior with parameter $\alpha$. For each $i^{th}$ word $x_{ij}$ in $D_j$, a topic $z_{ij} = k$ is drawn from $\theta_j$, and $x_{ij}$ is drawn from $\phi_k$.  The generating process for LDA is 
\begin{equation}
\theta_j \sim Dir(\alpha), \phi_k \sim Dir(\beta), z_{ij} = k \sim \phi_k.
\end{equation}

LDA has several implementations. There are two standard formulations, depending on how the Dirichlet priors are updated for the next iteration. The version used in the original paper and implementation uses variational Bayes, while many later works rely on Gibbs sampling \cite{TOT,li2006continuous,nallapati2007multiscale,wei2007dynamic,song2008non,jo2011web,PLDA,Wang-peacock}.  
The implementation of Gibbs sampling is less complicated and comparatively straightforward to derive.

Serial implementations of LDA do not scale well to processing of even moderately large corpora. Parallel LDA (PLDA) was developed by researchers at Google Beijing Research and Carnegie-Mellon University to address the processing of very large datasets \cite{PLDA}.  PLDA builds on a method called Approximate Distributed LDA (AD-LDA) \cite{ADLDA}. Instead of a probability mass function, topics are stored as the count of each word assigned to them. For example, if a word is generated by a topic fifteen times and there are sixty words generated by that topic in total, AD-LDA records '15', whereas LDA records '0.25'. This method utilizes data parallelism by dividing the set of documents across processes and iterating over the corpus using Gibbs sampling. Each process has a copy of the word counts.  Each process communicates any changes it makes to word assignment in its documents (and thus the resulting topic matrix) at the end of every iteration. During each iteration processes do not communicate, and thus operations that occur during an iteration are processed on stale intermediate values that are not globally accurate. As such, this approach is an approximation to serial Gibbs sampling. Experiments show this approximation converges in practice.  PLDA  demonstrates substantial speedup on large corpora over serial LDA implementations. 

Another parallel implementation, PLDA+, extends the implementation of PLDA and goes further by optimizing the algorithm using four strategies of data placement, pipeline processing, word bundling, and priority-based scheduling \cite{PLDAplus}. Data placement enables the pipeline to mask communication delays with further computation, working on one word bundle while communicating the results of another. These word bundles are chosen such that the computation time is long enough to mask communication, and arranged in a circular queue rather than statically assigned to processes. The queue and word bundles are managed by one set of processors while another set performs the Gibbs sampling, thus taking advantage of model parallelism. PLDA+ succeeds in masking communication with computation, and as a result has high scalability and performance to even PLDA, which is already fast. PLDA+ nears the theoretical maximum speedup for hundreds of processes and remains very high for all process counts tested.

\subsection*{Dynamic Topic Modeling}
One of the assumptions of LDA is that every document is equally important, but when evaluating documents over a long span of time this is problematic. For example, since language changes over time, the classification of a document written in 2000 should be based more on how it compares with documents written in the 1990s than in the 1900s. This problem can be partially sidestepped by considering blocks of time as separate collections, and performing LDA on each of them independently. This has the advantage of reducing the size of the corpus being used on any given task, which makes the method faster.  But, without further processing, it has the disadvantage that it loses the information about how a topic evolves over time.  CLDA addresses this problem directly.

DTM is one approach to the time dependency problem \cite{DTM}. Documents are sorted into discrete time steps, each containing a sizable corpus of its own. Each time step has its own topics, multiplying the size of the output by the number of time steps. We refer to the set of topics linked to each other over time as a \emph{global topic}, where its representation at a given time step is a \emph{local topic}. During each iteration, topics are updated by repeated inference on documents in their own time step, and also by consideration of the topic's form in the preceding time step.

DTM is effective in capturing the transformation of a global topic over time. It maintains the core strength of LDA while also allowing for variance across time periods to account for slowly changing language. However, there are some limitations.  There is no mechanism in DTM to capture the birth or death of topics. Also, the evolution model for the topics assumes the topics are recognizable from one year to the next. While a topic might gradually evolve to be unrecognizable from its original form, each individual jump must be smaller than the distance from that topic to the others in that time.

DTM also retains the parameterization of LDA to require as input the number of topics to be inferred. DTM adds further complication to this requirement, as the optimal number of topics may vary by time, which is not supported by the model.  The time series over the data segments enables the use of Gaussian models for the time dynamics.  However, the multinomial model of LDA and the Gaussian model for the time dynamics are non-conjugate, making posterior inference intractable, and an approximation must be used. Using such approximations at each iteration and splitting the time steps both negatively impact the convergence rate, so DTM often converges very slowly.

Investigation of dynamic topic modeling approaches that improves these weaknesses and increases performance is an active area of research \cite{timeline,diao2012finding,dubey2013nonparametric,chen2012dependent,he2009detecting,tu2010citation,xu2014author,lim2014bibliographic,dynamic-joint-sentiment,spatial-temporal-topic-model}. Table~\ref{tab:model_comparison} provides an overview of some of the most important developments and how they compare to CLDA. Topics over time (TOT) \cite{TOT} is a dynamic topic model that assumes continuous rather than discrete timestamps. iDTM \cite{timeline} is an extension of DTM that relaxes the assumption that the number of topics is fixed over time, therefore allowing for the birth and death of topics. This is achieved by combining the over-time updating from DTM with a hierarchical Dirichlet process (HDP) \cite{HDP}, which is a model built for nested data and which has been parallelized by \cite{parallelHDP}.

\begin{table*}[t]
\begin{footnotesize}
\begin{center}
\caption{\bf{Comparison of existing topic modeling approaches and CLDA} \label{tab:model_comparison}}
\begin{tabular}{l*{9}{c}}
\hline
& & & & & & parallel & & & parallel \\
& CLDA & LDA & PLDA+ & TOT & HDP & HDP & DTM & iDTM & DTM \\
\hline \hline
Reference & & \cite{LDA} & \cite{PLDAplus} & \cite{TOT} & \cite{HDP} & \cite{parallelHDP} & \cite{DTM} & \cite{timeline} & \cite{ParallelDTM} \\
Parallelized & $\checkmark$ & - & $\checkmark$ & - & - & $\checkmark$ & - & - & $\checkmark$ \\
Includes time component & $\checkmark$ & - & - & $\checkmark$ & ($\checkmark$)$^{\dagger}$ & ($\checkmark$)$^{\dagger}$ & $\checkmark$ & $\checkmark$ & $\checkmark$  \\
Evolution of topics & $\checkmark$ & - & - & $\checkmark$ & - & - & $\checkmark$ & $\checkmark$ & $\checkmark$ \\
Allows for birth/death of topics & $\checkmark$ & - & - & $\checkmark$ & - & - & - & $\checkmark$ & - \\
Unlimited number of segments & $\checkmark$ & - & - & $\checkmark$ & $\checkmark$ & $\checkmark$ & $\checkmark$ & $\checkmark$ & $\checkmark$ \\
Multiple segmentation options & $\checkmark$ & - & - & - & $\checkmark$ & $\checkmark$ & - & - & - \\
\hline
\multicolumn{10}{p{0.9\textwidth}}{\emph{Notes:}$^{\dagger}$ HDP was built for nested data. Similar to the modeling approach presented in this paper, HDP could be applied to time-segmented data to estimate changes in topics over time.} \\
\end{tabular}
\end{center}
\end{footnotesize}
\end{table*}

CLDA, described in the next section, utilizes parallel computing by applying LDA independently to each data segment and combining the results using clustering. 
Bhadury et al. \cite{ParallelDTM} devise a parallel method to address the normal complications with DTM's inference algorithm. Previous work relies on mean field approximations, which are costly to calculate.
Their work instead utilizes developments in stochastic Markov Chain Monte Carlo methods, a category which also includes Gibbs sampling. This allows them to utilize the more easily parallelized Gibbs sampling framework to estimate posterior likelihood, but is also faster in serial operation. Their results show dramatic speedup over the original DTM implementation, but the code is not available as of this writing for comparison to CLDA. The focus of our contribution is the expanded capabilities of CLDA over DTM rather than simply being a faster approximation, so we do not consider their work to compete with our own.

CLDA utilizes clustering in the final step of the method.  Because of its speed and available code, a parallel implementation of k-means developed in \cite{parallelk-means} was used for this project.  Inputs to the k-means method include the number of clusters, $K$, and an initial set of points.  
The data are classified to the nearest point using cosine similarity. 
There are weaknesses to this method.  There is an assumption that clusters should be roughly equally sized and that data in different clusters will be separated by considerable distance. The selection of $K$ can be problematic, and the algorithm is known to be sensitive to the initial starting points.  Because of these weaknesses, exploration of other clustering approaches is a topic of future work.


\section*{Method Description}\label{sec:method}
\begin{figure}[!ht]
\begin{center}
\includegraphics[width=0.4\textwidth]{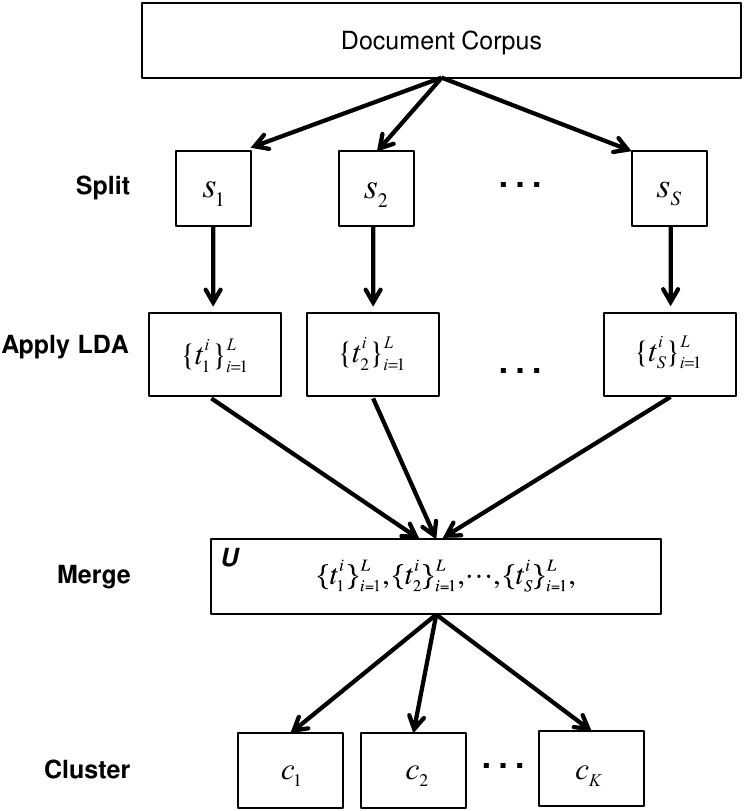}
\caption{\bf{Flowchart of Algorithm~\ref{fig:pseudocode}\label{fig:clda_flowchart}}}
\end{center}
\end{figure}

CLDA uses a data decomposition parallelization strategy.  The data are split into multiple segments and LDA is applied to estimate local topics in each segment in parallel.  Then, local topics are merged and clustering is used to calculate global topics on the merged local topics.  Algorithm~\ref{fig:pseudocode} provides the pseudocode and Figure~\ref{fig:clda_flowchart} provides the flowchart for CLDA. We describe the steps of CLDA in detail.

\setlength{\algomargin}{1em}
\begin{algorithm}[!ht]
    \SetKwInOut{Input}{Input}
    \SetKwInOut{Output}{Output}
\SetKwFunction{Split}{SPLIT}\SetKwFunction{ApplyLDA}{APPLY LDA}\SetKwFunction{Merge}{MERGE}\SetKwFunction{Cluster}{CLUSTER}
\Input{\: Number of segments $S$, number of local topics $L$, number of global topics $K$}
\Output{\: $L \times S$ local topics, grouped into $K$ clusters}
\Split text corpus into S segments \;
\For(\tcp*[h]{runs in parallel}){all segments $s\in \{1,...,S\}$}{
\ApplyLDA to estimate local topics $\{t^i_s\}_{i=1}^L $ \;
}
\Merge $(\{t^i_s\}_{i=1}^L $ for $s\in \{1,...,S\}) \rightarrow U$ ; \tcp*[h]{runs in parallel, see Algorithm~\ref{fig:pseudocode_merging}} \;
\Cluster $U$ into $K$ global topics \;
\caption{\bf{Pseudocode for Clustered Latent Dirichlet Allocation (CLDA)\label{fig:pseudocode}}}
\end{algorithm}

\subsection*{\emph{\textsc{Step 1: Split} text corpus into $S$ segments}}
First the corpus is divided into $S$ segments on which LDA will be applied. In our application we divide the data according to the naturally occurring disjoint segments of time steps (yearly data for each corpus) and journal publication (for the Pubmed corpus). Other applications might divide data by geographical location or data source. The division of the overall corpus into individual segments can be performed as a serial task or in parallel. The vocabulary is distributed to all tasks prior to the LDA computation, and the remaining data manipulation before each LDA executes independently on the individual, smaller, segments. The smaller the segments, the more efficiently this approach can utilize parallel resources, as segments can be processed in parallel to each other and each individual segment will be processed more quickly due to the reduced data size. However, segments that are too small risk compromising result quality, as discussed later.

LDA requires that the number of estimated topics, $L$, be selected {\em a priori}. The number of local topics $L$ can be larger or smaller than the number of global topics $K$.  We have found that often better results are obtained when the number of local topics $L$ is larger than what may be expected for global topics. The larger number of topics at the local level allows for small topics to be discovered, and for greater breadth of a topic that is unusually well represented. If many topics represent the same subject at any given segment, these are clustered together. 

In this paper we describe the case where $L$ is constant for each segment.  Extensions are possible where a different $L$ is set for each segment, either for domain-specific reasons or after determining the locally optimal number of topics through cross-validation.

\subsection*{\emph{\textsc{Step 2: Apply LDA} to estimate local topics}}
In the second step the documents in each segment are analyzed with LDA. LDA can run concurrently on separate processors (or groups of processors, if using parallel implementations of LDA such as PLDA in our experiments) for nearly perfect parallelism. This step results in a collection of $L$ topics $\{t^i_s\}_{i=1}^L $ at each segment $s\in \{1,...,S\}$, for a total of $S\cdot L$ local topics (whose merged union is denoted by $U$ in Algorithm \ref{fig:pseudocode}) that are clustered in the next stage.

\subsection*{\emph{\textsc{Step 3: Merge} local topics}}
The third step is to merge the emitted topics into a single collection $U$ which can be input to the clustering method. At the conceptual level this requires concatenating the emitted topics into a single list, but in practice this step is more involved. The individual outputs $\{t^i_s\}_{i=1}^L $ have indexing entries that must be removed before they can be concatenated. The entries are then re-indexed to match the input requirements of the chosen implementation of k-means (here \cite{parallelk-means}). It is also necessary to ensure that the generated topics are comparable. LDA acts on a vocabulary consisting of everything that appears in its source documents, and produces topics with a value for each element in the vocabulary. If a word appears in one document collection but not another, the resulting topics are not directly comparable. As such, if any of the segments does not contain the full vocabulary, it is necessary at this stage to add the missing entries to the topics, as shown in Algorithm~\ref{fig:pseudocode_merging}.  The entries are added with zero contribution to the topic. Depending on application, it may be valuable to instead set these values to some small value $\epsilon$, or set $\{t^i_s(w)\}_{i=1}^L \Leftarrow \{t^i_s(w)\}_{i=1}^L + \epsilon$ for all topic entries. Either adjustment can be performed in this stage with minimal performance cost, but our implementation leaves the topics untouched beyond the addition of zeros for missing words.

\begin{algorithm}[!ht]

    \SetKwInOut{Input}{Input}
    \SetKwInOut{Output}{Output}
\Input{\: Number of segments $S$, full vocabulary $W$, local vocabularies $W_s$, local topics $\{t^i_s\}_{i=1}^L$ }
\Output{\: Topic set $U$, containing all local topics in a shared vocabulary space}
\For(\tcp*[h]{runs in parallel}){all segments $s\in \{1,...,S\}$}{
\For{all words $w \in W$}{
\If{$w \notin W_s$}{
add $w$ to $W_s$ \;
\For{all local topics $\{t^i_s\}_{i=1}^L$}{
$t^i_s(w) \leftarrow 0$ \;
}
}
}
}
$U \leftarrow \bigcup_{s=1}^S \{t^i_s\}_{i=1}^L$ \;
\caption{\bf{Pseudocode for MERGE step\label{fig:pseudocode_merging}}}
\end{algorithm}


In addition to ensuring the local topics are comparable in dimension, they must be comparable in scale. Some LDA implementations, including PLDA, provide varying magnitudes for topic vectors based on their occurrence in the data.  The goal of CLDA is to cluster the local topics based on the meaning, and we assume that all local topics are equally weighted. As such, the topics are normalized before clustering them. This operation is straightforward and has no dependence on other topics or other segments, and can thus be done independently before the merge, or all at once afterwards. Our implementation performs this normalization before the merge, but there is no difference in the results either way.

\subsection*{\emph{\textsc{Step 4: Cluster} local topics}}
The fourth step is to combine local topics into global topics. The k-means clustering requires that the number of global topics, which we denote as $K$, is set {\em a priori}. 
 
In the extreme cases, $K=1$ defines a single cluster containing all local topics, and $K=(S \cdot L)$ defines a cluster for every topic individually. If $K > L$, not every global topic will have a representation at each segment, which means global topics will disappear and/or reappear. If $K \leq L$, topics \emph{may} disappear and reappear at individual segments, depending on the results of the clustering. We consider this to be an advantage of the method over alternative implementations, such as DTM, which assumes that topics are universally represented over the entire time period.  

k-means clustering is sensitive to its initialization. In CLDA, this may result in different topics. 
There are ways to evaluate the output of k-means across different initial values, including inter-class sum of squares, but generating initial values that are sufficiently different from each other is still a data-dependent challenge. Running LDA on the entire corpus provides a set of topics that make an intuitive set of initial values for clustering, regardless of data properties. While this can be done concurrently with running LDA on individual segments, it takes longer to complete than individual segments due to the larger corpus. To avoid the performance impact, we can use fewer iterations on the full corpus than we do on the local segments. Alternatively, instead of running LDA on the entire corpus, choose $K$ random topics from $U$ as the initial values. In our implementation, we support both options, and we report results from both approaches in our results. 

\subsection*{\emph{\textsc{Step 5: Output} local topics and global topic assignments}}
Once clustering is complete there are two important outputs. The first is the centroids themselves, each usable as a topic, and the second is the assignment of the original topics to their corresponding global clusters. The centroids are useful on their own, and provide summary information DTM does not. DTM does not provide a general vision of a given dynamic topic, only its local topics at each time step. CLDA provides both a segment-agnostic version of a topic and a varying number of local topics at each segment, including potentially none at all, which would indicate that the topic was not meaningfully present in that segment.

\section*{Experimental Validation}\label{sec:validation}


We evaluate our method on three different data sets, which are summarized in Table~\ref{tab:data_summary}. Our first data set is a collection of all NIPS papers from 1987 to 2003 \cite{chechik2007eec}, which we selected because it is a widely used data source for evaluating the quality and performance of topic models. The NIPS data contains 2,484 documents (about 150 documents per time segment), 14,036 unique words, and 3,280,697 tokens. Our second data source is a collection of abstracts from published articles in computer science provided by Elsevier and pre-processed using the open source HPCC Systems platform by LexisNexis. This data set covers the same number of time segments as the NIPS data, but includes a much larger number of documents ($N=533,588$) with about 31,000 documents per time step. The computer science abstracts data contains 17,998 unique words and 46,446,184 tokens after our pre-processing. We remove stopwords using the NLTK stopword definition, as well as any word that appears fewer than 100 times or in fewer than 10 unique documents. We also remove single-character words and words containing only symbols or numbers. This corpus is much broader in scope and hence requires a greater number of topics to describe the documents. Our third corpus is a forty year collection of article abstracts from PubMed for the time period from 1976--2015, which contains 4,025,976 documents after we removed non-English abstracts and all articles published in journals with less than 10,000 total number of articles. The PubMed corpus contains 4,025,976 documents, 69,742 unique words and 386,847,695 tokens after being pre-processed the same way as the computer science abstracts data. We use this dataset to demonstrate the scalability of the approach, and the use of alternate segmentations of the same data. \emph{Having a scalable implementation capable of producing dynamic topics is extremely important in the analysis of biomedical literature to accelerate scientific discovery.} Examples include analysis of scientific trends \cite{spreckelsen2016my,li2015bibliometric} and topic modeling based hypothesis generation \cite{sybrandt2017moliere} which makes the ABC model \cite{swanson1997interactive,smalheiser1998using} more interpretable. Non-scalable topic modeling often represents a major bottleneck in qualitative scientific literature analysis \cite{sybrandt2018abstracts}.

\begin{table}[!ht]\centering
\caption{\bf{Overview of data used for evaluation\label{tab:data_summary}}}
\begin{tabular}{l c c c c }
\hline 
& NIPS  & Computer Science & PubMed \\
& & Abstracts \\
\hline
Time period & 1987--2003 & 1996--2012 & 1976--2015 \\
No. of time segments & 17  & 17 & 40   \\
No. of journal segments & N/A & N/A & 208 \\
No. of documents & 2,484 & 533,588 & 4,025,976 \\
Vocabulary size & 14,036 & 17,998 & 69,742 \\
Total word tokens & 3,280,697 & 46,446,184 & 386,847,695 \\
\hline
\end{tabular}
\end{table}

All experiments were performed on Clemson University's Palmetto Cluster, using Intel Xeon Gold 6148 nodes with 40 cores, 748 GB of RAM, and 56 Gbps interconnects. Our implementation of CLDA for these experiments utilizes PLDA \cite{PLDAplus} for the LDA stage, and k-means \cite{parallelk-means} for the clustering stage. The data manipulation code connecting the stages is written in Python 2.7 and Python 3.6, and can be found here: \cite{CLDAimplementation}. Future work will explore alternative implementations of the various CLDA components.

\subsection*{Performance}
We first compare CLDA's runtime with the DTM implementation by Blei and Gerrish \cite{DTMcode} on the computer science abstract data. We execute the DTM model on 20 topics, and the CLDA model using 20 global topics and 20 local topics per segment. Blei and Gerrish's DTM implementation is not parallelized, and PLDA does not run in serial, so we are unable to perfectly match the resources used. All timing experiments were run on the same hardware, only varying processor count.

The results shown in Table~\ref{tab:runtimes} demonstrate that the algorithm is orders of magnitude faster than the original implementation of DTM. This is unsurprising; the primary operation of consequence is the LDA phase of the algorithm, where our implementation utilizes the highly optimized PLDA. The other operations largely consist of data manipulation to normalize or rotate files, and the clustering step. However, the clustering input is small compared to the size of the input data, so in our experiments k-means converged in seconds.

\begin{table}[!ht]\centering
\caption{\bf{Runtime results on computer science abstracts 
\label{tab:runtimes}}}
\begin{footnotesize}
\begin{tabular}{lcccc}
\hline
 & \# of & Iterations & Walltime & Walltime \\
 & Cores & Iterations & (max) &  (sum) \\
\hline
DTM  & 1 & 100 & 3497 & 3497 \\
PLDA & 12 & 1,000 & 41 & 41 \\
CLDA & 12 & 1,000 & 3 & 33 \\
\hline
\end{tabular}
\end{footnotesize}
\end{table}

We next evaluate CLDA on the PubMed data, which is an order of magnitude larger than the computer science abstracts data. Additionally, we have enough data for enough different journals to examine the PubMed data using each of time segmentation and journal segmentation. \emph{The ability to examine alternative data segmentations is a key contribution of CLDA.} 
The LDA phase of the algorithm concluded in 22 minutes on this dataset using 1,000 iterations and 12 cores per segment, a total of 480 cores overall. This performance takes advantage of multiple levels of parallelism, as CLDA is able to process each segment simultaneously and PLDA can leverage distributed computing within a segment. DTM applied to the same data with the same number of iterations would take approximately 29 weeks to complete given our earlier findings. 

\subsection*{Quality}
In order to be useful, the topics produced by the algorithm must be either very similar to those produced by DTM, or superior to them. Measuring the quality of a topic model is an open question, but a standard approximation is the perplexity metric. This metric evaluates how likely the topic model is to generate a set of provided documents. A lower perplexity indicates a model more closely fits the documents. As perplexity is a function of probabilities rather than direct model parameters, it can be used to compare different models over the same input. 

Perplexity is calculated using

\begin{equation}
 \text{perplexity} = \exp{\left(-\frac{\sum\limits_{d \in D}\; {\sum\limits_{w \in d}{\log{P(w|d)}}}}{\sum\limits_{d \in D}{N_d}}\right)}
 \label{eq:perplexity}
\end{equation}

where $d$ denotes a document in the corpus $D$, $w$ denotes a word, and $N_d$ denotes the number of tokens. We use a hold-out set to evaluate perplexity, executing the model on 80\% of the data and testing it on the remaining 20\%. To evaluate the probability of generating a word $P(w|d)$, it is necessary to generate topic mixtures for held-out documents. We use the code provided with PLDA for this task \cite{PLDA}, but a more thorough study of this problem can be found in Wallach et al. \cite{wallach2009evaluation}.

Note that in the case of CLDA, documents can be evaluated in the context of both local topics and global topics. We believe that evaluating document level perplexity using the local topics better represents the power of CLDA, and is comparable to how perplexity must be evaluated on DTM, since DTM lacks global topics. However, we have included the results for using global topics for completeness reasons.

Table~\ref{tab:perplexity} provides perplexity results for CLDA, DTM, and PLDA estimated on the full computer science abstract data. The DTM model was executed for 58 hours using 20 topics, while the CLDA models were executed in 10 minutes using $K=20$ global topics and $L=20$ local topics. PLDA estimated on the full data using 20 topics completed in 17 minutes. The results show that CLDA has a comparable perplexity to DTM and PLDA for this data set. 

\begin{table}[!ht]\centering
\caption{\bf{Perplexity results on computer science abstracts using 10-fold cross validation \label{tab:perplexity}}}
\begin{tabular}{lccc}
\hline
 & DTM & CLDA & PLDA \\
\hline
Perplexity & 1,950 & 1871 & 1885 \\
\hline
\end{tabular}
\end{table}

\subsection*{Similarity}
The previous results indicate that our system is both very fast and has competitive perplexity to other methods. We wish to know how similar the generated topics are to those generated by DTM or LDA.

Topics are probability mass functions represented by vectors, but this is not how humans interpret them \cite{ReadingTeaLeaves,Towne-similarity}. Rather than look holistically at the entire vector, a human will examine the most heavily weighted words in a topic; for example, the top five. These words will provide insight as to the conceptual meaning of a topic. In order to compare the insights gleaned from a set of topics, we thus need to compare what a human compares; the words most strongly tied to a topic. For a word-wise comparison of topics as sets of important words we will use the S{\o}rensen-Dice coefficient 
\begin{equation}
S(A,B) = \frac{2*|A\cap B|}{|A|+|B|}
\label{Dice}
\end{equation}
and the Jaccard index
\begin{equation}
J(A,B) = \frac{|A\cap B|}{|A\cup B|}
\label{Jaccard}
\end{equation}
where A and B are sets; in our context, A and B are representative sets of two topics being compared. Specifically, we use the top 20 most commonly generated words in a topic as its representative set. Chang et al. \cite{ReadingTeaLeaves} used the top 5 words as the core of a topic for their intruder experiment, but they were using humans to detect outliers instead of searching for broad similarity. We chose this value as it is low enough to be human-readable, but high enough to dampen the impact of minor value differences on ordering. However, this value is still arbitrary. Future work will explore other means of transforming topics into sets.

We compared the systems using both measures. We compared the global topics to each other by comparing their means. For our system, these are emitted by clustering, but for DTM we averaged the local topics together. Both the S{\o}rensen-Dice coefficient and the Jaccard index compare single sets to each other. Comparing the outputs of DTM and CLDA requires assigning a one-to-one matching between the two collections. If the topics generated by DTM and CLDA both include a topic describing the same concepts, these two topics will match more closely than they match other topics. If this is not the case, then the topics are not similar and a low value will be obtained regardless of the optimality of matching. Our experiment utilizes this assumption by greedily matching the pair of unassigned topics that are closest to each other under the Jaccard index out of all possible pairings, repeating the process until all topics are assigned. The Jaccard index and S{\o}rensen-Dice coefficient are calculated for each match. 

\begin{figure*}[!ht]
\begin{center}
\includegraphics[width=\textwidth]{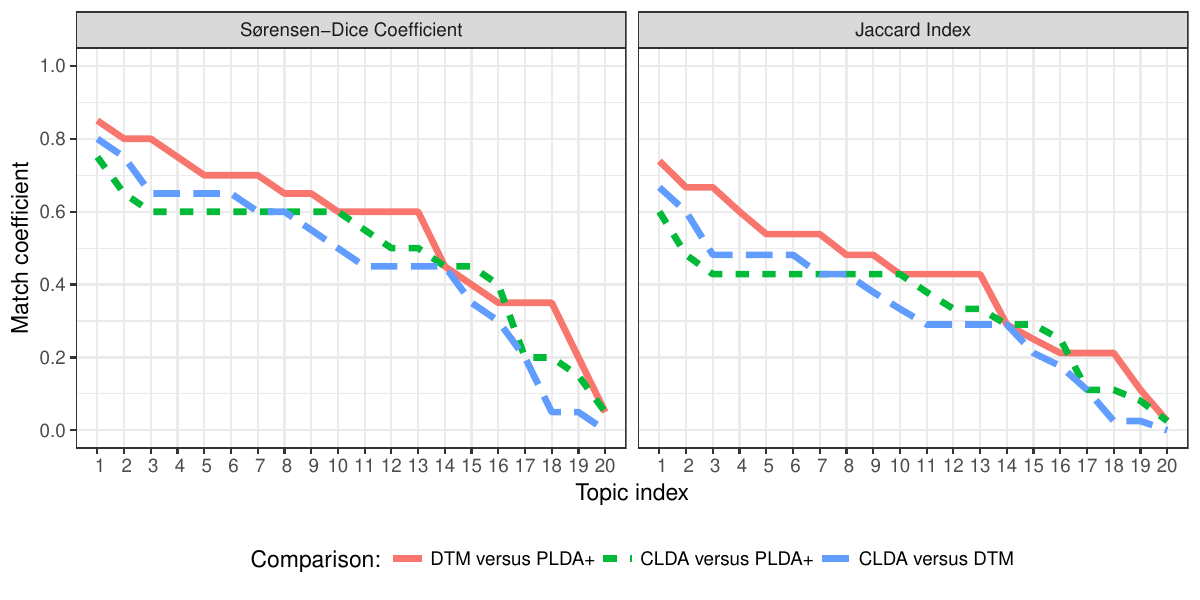}
\caption{Similarity of global topic centroids between DTM (estimated with 20 topics), PLDA (estimated with 20 topics), and CLDA (estimated with 20 global topics and 20 local topics) applied to the computer science abstracts data as measured by S{\o}rensen-Dice Coefficient and Jaccard Index. \label{fig:comparison_CLDA_DTM_LDA}}
\end{center}
\end{figure*}

The values of the global topic matches are shown in Figure~\ref{fig:comparison_CLDA_DTM_LDA} sorted from best to worst. We compare CLDA's global topics to those of both PLDA and DTM, and also compare PLDA to DTM directly as a reference point. 
We find that these comparisons all produce similar results; the global topics from each approach are all roughly the same distance from one another. 

\subsection*{Multiple segmentations}
Unlike other topic models, CLDA can be used on varying data segmentations rather than just one. In particular, we examine the PubMed data set segmented by each of time and journal. Note that while the full PubMed corpus contains articles from journals with only a comparatively small number of articles each, we have trimmed our corpus to only include abstracts from journals with at least 10,000 documents in the collection. This way, we ensure that each journal segment contains at least 10,000 documents, rather than having many segments with only a few documents.

We compare the results of these two segmentations on several axes. We examine their runtime and perplexity, using the methods described in our above experiments. The runtime and perplexity results are summarized in Table~\ref{tab:pubmed}.

\begin{table}[!ht]\centering
\caption{\bf{Runtime and Perplexity Results on PubMed corpus 
\label{tab:pubmed}}}
\begin{tabular}{lccc}
\hline
 & PLDA & CLDA (years) & CLDA (journals) \\
\hline
Walltime (minutes, max) & 121 & 9 & 11 \\
Walltime (minutes, sum) & 121 & 143 & 308 \\
\hline
Perplexity & 2657 & 2356 & 1392 \\
\hline
\end{tabular}
\end{table}

We find that CLDA allows for much better exploitation of parallel resources, with a total walltime an order of magnitude faster than that of PLDA. However, we do find that there is some cost to total resource utilization, especially when dealing with a large number of segments. We hypothesize that this is due to static initialization times, since the effect appears greatly magnified in the journal segmentation, which has 208 segments compared to the 40 of the year segmentation. Interestingly, this did not occur in the computer science abstract data, which had only 17 segments.

We also find that the global topics of the journal segmentation differ from those of the year segmentation. In particular, the topics generated by PLDA are substantially closer to those of the year segmentation than they are to those of the journal segmentation.  We also find that while the perplexity of the year segmentation is better than that of PLDA, the perplexity of the journal segmentation is vastly superior to that; we hypothesize that CLDA is taking full advantage of the smaller segments.

\subsection*{Global and local topic dynamics}
LDA can be used to capture change in topic proportions over time, by executing the model over a whole corpus and then evaluating segments of it. This does not capture any change in topic language over time, forcing each segment to use the same topics. DTM relaxes this constraint by allowing the topics to vary over time. DTM produces both a version of each topic at each segment, as well as the relative proportion of each topic at each segment, demonstrating how both language and representation change over time \cite{DTM}. However, DTM fixes the number of topics across time, with each overall topic having one representative per segment. CLDA relaxes this further, allowing a global topic to have any number of local representatives at each segment, including zero. In addition to allowing for topics to branch out, better fitting their local data, this also allows for global topics to appear and disappear entirely.

The strength of DTM is the variation of topics over time, taking on forms better suited to their local data while remaining tied together by a common theme. Blei et al. \cite{DTM} demonstrate this by examining the changing form of a topic at several time steps, as well as their changing proportions over time. CLDA produces output to provide this same type of insight into a corpus.

\begin{figure*}[!ht]
\begin{center}
\includegraphics[width=0.49\textwidth]{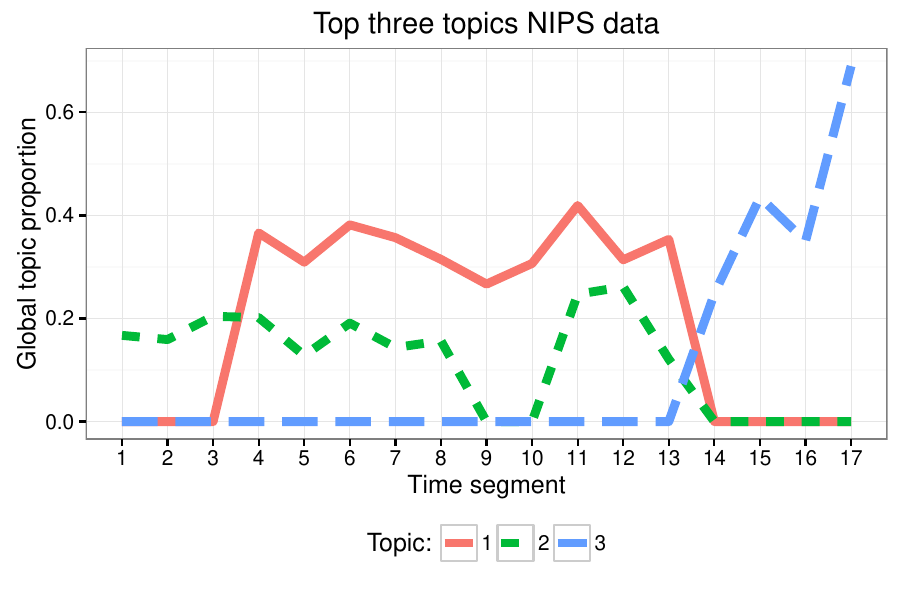} 
\includegraphics[width=0.49\textwidth]{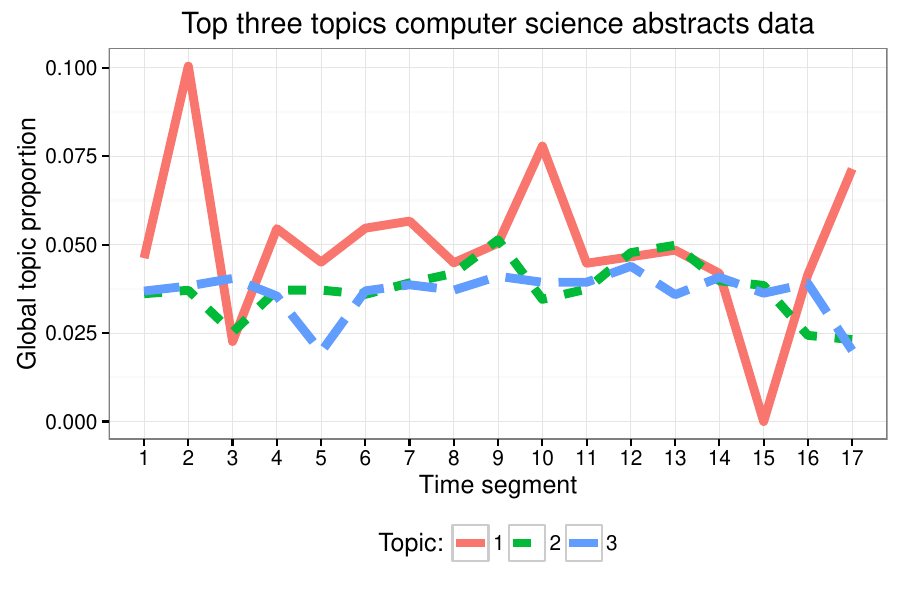}
\caption{Evolution of three largest global topics for the NIPS data (left panel) and computer science abstracts data (right panel).\label{fig:topic_evolution_abstracts}}
\end{center}
\end{figure*}

We show the changing topic proportions for selected topics in both the NIPS data and computer science abstract data in Figure~\ref{fig:topic_evolution_abstracts}. Like DTM, CLDA provides insight into the rising and falling predominance of various topics in a corpus. Unlike DTM, CLDA global topics need not be composed of exactly one topic at each segment. 

\begin{figure*}[!ht]
\begin{center}
\includegraphics[width=\textwidth]{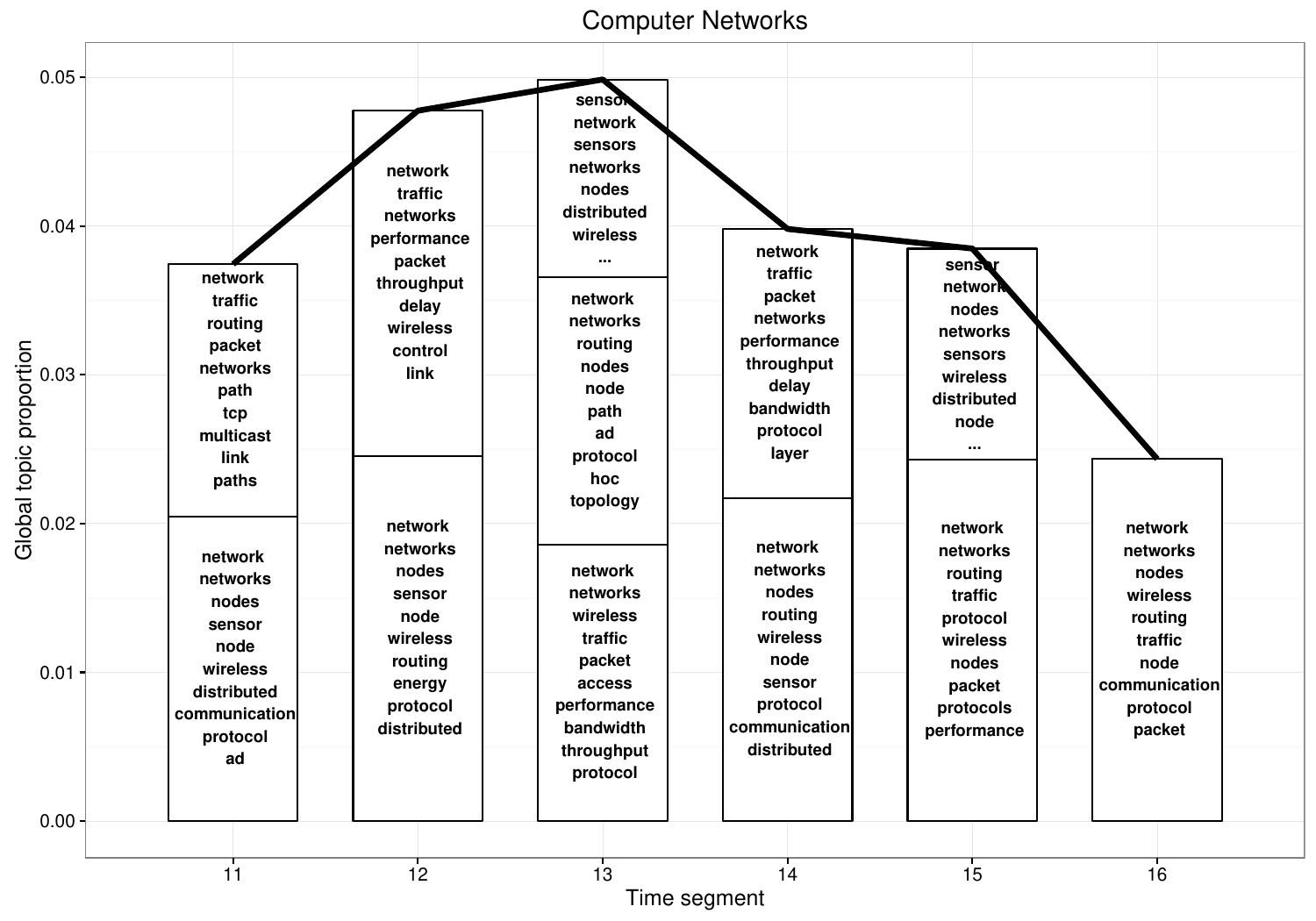}
\caption{Local topics for selected time segments corresponding to global topic ``Computer Networks'' from the computer science abstracts data using 62 global topics and 50 local topics in each segment. Each bar lists the top words in each local topic. The height of each bar corresponds to the proportion a local topic contributes to the global topic.\label{fig:selected_topics_with_word}}
\end{center}
\end{figure*}

Figure~\ref{fig:selected_topics_with_word} shows how a changing number of local topics represent a global topic we identify as ``Computer Networks'' for six selected time segments from the computer science abstract data. While these topics are all clustered together, they represent distinct ideas within the overall concept of ``Computer Networks''. One may focus on software defined networking, while another may focus on the communication between remote sensors. While this distinction is useful to examine, treating these as fully separate topics does not produce an accurate picture of how prevalent computer networks research is in the corpus as a whole. Clustering these topics together provides both the global insight of overall representation and local insight into a research area's subdomains.

\section*{Conclusion}\label{sec:conclusion}

We have constructed and evaluated CLDA, a method for analyzing topic dynamics in text data. This algorithm leverages existing parallel components to increase speed and facilitate the use of large corpora. It begins by discretizing the data into disjoint segments, and applying Latent Dirichlet Allocation on each segment in parallel. The resulting local topics are merged and then k-means clustering is applied, producing a number of global topics. Each global topic is composed of a number of local topics in each segment, and provides a summary of the cohesive theme across segments. Our system is built using PLDA \cite{PLDA} and parallel k-means clustering \cite{parallelk-means}. 

We find that our system performs faster than the original implementation of DTM by two orders of magnitude. CLDA also has perplexity comparable to that of DTM and PLDA. The topics generated by CLDA, DTM, and PLDA are all broadly similar to each other. CLDA shows a more detailed composition of local topics than is possible with DTM, and enables global topics to emerge and disappear over the time span. We also demonstrate CLDA's application to examining multiple segmentations of the same data, a capability not possessed by other models we are aware of. Taken together, these results show that CLDA is a promising approach for modeling dynamics in topics estimated from textual data. The implementation of CLDA is available at \cite{CLDAimplementation}.

Future work may explore alternative clustering approaches, such as using hierarchical clustering instead of k-means. CLDA can also be applied to data segmented by factors other than time or journal, such as author or location. Also of interest is the effect of modifying the segmentation approach, either by changing the coarseness of the segments or by exploring non-disjoint segmentations of the data. Instead of segmenting the data, CLDA can also be used to aggregate topics across multiple runs of LDA on the same data but with different parameters, which would allow identifying topics that are stable with respect to model parameterization. 
\pagebreak
\bibliographystyle{IEEEtran}
\bibliography{references,refs2}

\end{document}